\documentclass[conference]{IEEEtran}
\IEEEoverridecommandlockouts
\usepackage{cite}
\usepackage{amsmath,amssymb,amsfonts}
\usepackage{listings} 
\usepackage{algorithmic}
\usepackage{graphicx}
\usepackage{textcomp}
\usepackage{xcolor}
\usepackage{tabularray}
\usepackage{hyperref}
\usepackage{color}
\usepackage{array}
\usepackage{multirow}
\usepackage{ragged2e}
\def\BibTeX{{\rm B\kern-.05em{\sc i\kern-.025em b}\kern-.08em
    T\kern-.1667em\lower.7ex\hbox{E}\kern-.125emX}}

    \makeatletter
    \newcommand{\linebreakand}{%
      \end{@IEEEauthorhalign}
      \hfill\mbox{}\par
      \mbox{}\hfill\begin{@IEEEauthorhalign}
    }
    \makeatother
    
\begin{document}

\title{A Reconfigurable Approximate Computing RISC-V Platform for Fault-Tolerant Applications}

\author{
\centering
\IEEEauthorblockN{A. Delavari, F. Ghoreishy, H. S. Shahhoseini, S. Mirzakuchaki}
\IEEEauthorblockA{\textit{School of Electrical Engineering} \\
\textit{Iran University of Science and Technology}\\
Tehran, Iran \\
\{arvin\_delavari, faraz\_ghoreishy\}@elec.iust.ac.ir, \{shahhoseini, m\_kuchaki\}@iust.ac.ir}
}

\maketitle

\begin{abstract}
    The demand for energy-efficient and high-performance embedded systems drives the evolution of new hardware architectures, including concepts like approximate computing. This paper presents a novel reconfigurable embedded platform named ``phoeniX'', using the standard RISC-V ISA, maximizing energy efficiency while maintaining acceptable application-level accuracy. 
    The platform enables the integration of approximate circuits at the core level with diverse structures, accuracies, and timings without requiring modifications to the core, particularly in the control logic.
    The platform introduces novel control features, allowing configurable trade-offs between accuracy and energy consumption based on specific application requirements. 
    To evaluate the effectiveness of the platform, experiments were conducted on a set of applications, such as image processing and Dhrystone benchmark.
    The core with its original execution engine, occupies 0.024mm² of area, with average power consumption of 4.23mW at 1.1V operating voltage, average energy-efficiency of 7.85pJ per operation at 620MHz frequency in 45nm CMOS technology. The configurable platform with a highly optimized 3-stage pipelined RV32I(E)M architecture, possesses a DMIPS/MHz of 1.89, and a CPI of 1.13, showcasing remarkable capabilities for an embedded processor. 
\end{abstract}

\begin{IEEEkeywords}
Approximate Computing, Low Power Design, RISC-V, Embedded Processor, Energy-Efficient Computation, Very Large Scale Integration
\end{IEEEkeywords}

\section{Introduction}\label{introduction-section}

    Approximate computing is an emerging paradigm for energy-efficient and high-performance designs. It includes various sets of computation techniques that return a possibly inaccurate result rather than a guaranteed accurate result. This approach is useful in applications where an approximation is sufficient for its purpose. 
    Approximation techniques are employed in applications such as image processing \cite{design-of-approximate-adders-and-multipliers-for-error-tolerant-image-processing}, neural networks \cite{neural-networks}, machine learning and PIM methods \cite{in-memory-machine-learning} due to their inherent capabilities of adhering inaccurate results. By utilizing approximation, these applications can handle large computational workloads, maintaining acceptable accuracy.

    The proposed platform is crafted in a way to be a global and high-quality foundation for integration of approximate computing techniques, with the RISC-V architecture. The general idea of the scheme is integration of approximate arithmetic circuits into a highly optimized 3-stage pipelined processor with minimum internal fragmentation \cite{modern-processor-design}, constructing an embedded core, suitable for error-tolerant applications. 

    The modular architecture of the platform enables designers to add and test approximate arithmetic circuits on the core level, without any need for changes in other parts of the processor such as control logic and etc. With this capability, designers can enhance performance by adding new features, and develop different architectural techniques. This project has been open-sourced\footnote{\url{https://github.com/phoeniX-Digital-Design/phoeniX}} with technical documentation, helping designers to fully understand the structure of the platform.

    In this work, a processor is introduced and described as a novel approximate computing development platform using standard RISC-V instruction set architecture and conventions. The major contributions of this paper are as follows:
    \begin{itemize}
    
        \item Introduction of the phoeniX core, a modular and reconfigurable, 32-bit optimized 3-stage pipelined RV32I(E)M embedded processor platform for approximate computing applications with specialized features, designed in Verilog HDL.
        
        \item Introduction of a customized control logic for accurate/approximate execution units with error-configurablity feature, using RISC-V Control Status Registers (CSRs), and dynamic internal circuit switching using CSRs for all execution units integrated within the core.
        
        \item Design and implementation of an accuracy controllable, very fast and low-power Carry Select Adder circuit as the default (demo) adder circuit of the platform.
        
        \item Design and implementation of an error configurable, high-speed and low-power approximate multiplier circuit as the default (demo) multiplier circuit of the platform.
        
        \item Design and implementation of a dynamic accuracy controllable non-restoring divider unit, with accurate division capability at the base error level.
    
    \end{itemize}

    The rest of the paper is organized as follows: In Section \ref{related-works-section} investigations on related works are discussed. In section \ref{architecture-section} the architecture of the platform is described in details. In Section \ref{execution-engine-section}, the execution engine of the core is explained and Section \ref{application-section} is considered for experimental results on an image processing application. Hardware implementation results and comparison are included in Section \ref{synthesis-section} and finally, the paper is concluded in Section \ref{conclusion-section}.

    \begin{figure*}[htbp]
    \centering
    \includegraphics[width=\linewidth]{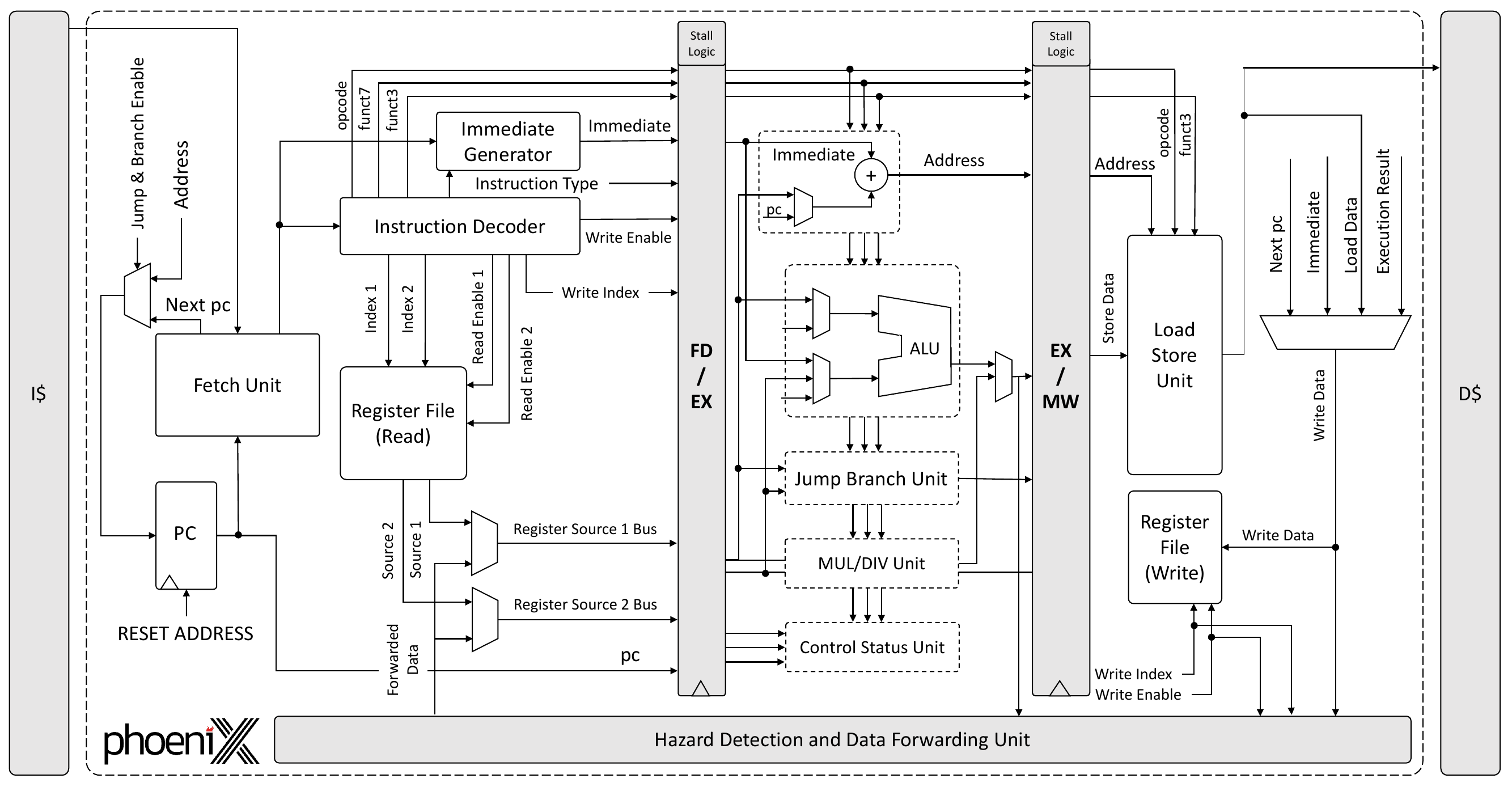}
    \caption{The phoeniX RISC-V processor block diagram}
    \label{block-diagram-figure}
    \vspace{-3.5mm}
    \end{figure*}

\section{Related Works}\label{related-works-section}
    Over the last few years, approximate computing has enhanced the energy efficiency of many applications. There has been recent exploration into the utilization of RISC-V in combination with approximate computing methods. Software adaptation has resulted in the proposal of an ISA extension, along with a control mechanism for multi-level accuracy, which is discussed in \cite{Approxrisc}. Software approximation methods can be effective in certain cases, it is important to recognize that approximate hardware techniques are often highlighted and evaluated independently, and are typically implemented in customized application-specific settings.
    AxPIKE, an ISA simulator that enables incorporation of hardware approximation at the instruction level and assess their impact on result, is introduced at \cite{AxPIKE-paper}. By implanting approximation in multiplication and memory access instructions of various applications, the authors demonstrate how the resulting statistics are used to evaluate quality trade-offs with respect to energy. The study in \cite{risk-5} by Felzmann et al. characterized an extension for a RISC-V architecture that orchestrates diverse circuit-level approximation. The authors aimed to connect software and hardware approximation methods for better outcomes.
    
    In \cite{riscv-multiplier}, a customized core based on RI5CY \cite{ri5cy-paper} is implemented, by integrating an approximate multiplier within. 
    MARLIN \cite{marlin-paper} is also a framework that enables the deployment of approximate neural networks (NNs) on energy-constrained devices like microcontrollers. It integrates a customizable multiplier architecture with runtime selection of 256 approximation levels into a PULP microcontroller \cite{pulp-paper-2}, allowing for energy-efficient execution of NNs.
    
    Investigations were conducted recently on advantages of a Multi-Processor System-on-Chip (MPSoC) incorporating a blend of both exact and approximate cores. The integration of approximate computing in company with accurate computation capabilities in an MPSoC holds potential for leveraging the benefits offered by approximation, while maintaining the precision provided by exact computing. In \cite{axe}, a multi-core system with exact and approximate cores named AxE is introduced. AxE is a heterogeneous RISC-V MPSoC with exact and approximate cores using PicoRV32
    \cite{pico}, that allows exploring hardware approximation for any application by using separate software instructions for approximate and exact computation.
    
    Our approach presents a different perspective on the integration of RISC-V processors and approximate computing. The proposed design incorporates a single RV32I(E)M core where the output of each execution unit switches between accurate and approximate circuits by user's request. Parameters are included within the RTL source in order to switch between I and E extension, and enabling M extension of RISC-V standard ISA. Each execution unit of the platform can host four arithmetic circuits and is capable of switching between these circuits. Also, support for error control is provided if applicable in these modules.
    
    This method combines the precision of accurate circuits with the efficiency and potential energy savings offered by approximate circuits. While the MPSoC approach presented in \cite{axe} provides accurate and approximate computation in separate cores, the proposed design takes the benefits of putting approximation alongside exact computation capabilities in one standard core, with maintaining dynamic configurability features. This will lead to lower occupied area, energy saving, less complexity in software development and performance boost in a single core. Alternatively, it can be also placed back in an MPSoC platform where each node consisting of this core can use different execution units while benefiting from a similar control logic which is applicable in both hetero- and homogeneous topologies.

\section{Architecture}\label{architecture-section}
    The proposed processor, is a 32-bit in-order scalar core which the high-level datapath is shown in Fig.~\ref{block-diagram-figure}. It features a modular and extensive design, which is highly beneficial for designers, as it facilitates testing and research on computer architecture techniques. A key aspect of this processor is the removal of the traditional ``Control Unit'' found in most RISC processors. Instead, each module in the processor can generate its own control signals based on the fields of the instruction it receives after the decode stage. This micro-architecture is regarded as distributed control logic in pipeline processors \cite{modern-processor-design}, and it helps us with the modularity of the core.

\begin{figure}[htbp]
    \centering
    \includegraphics[width=\linewidth]{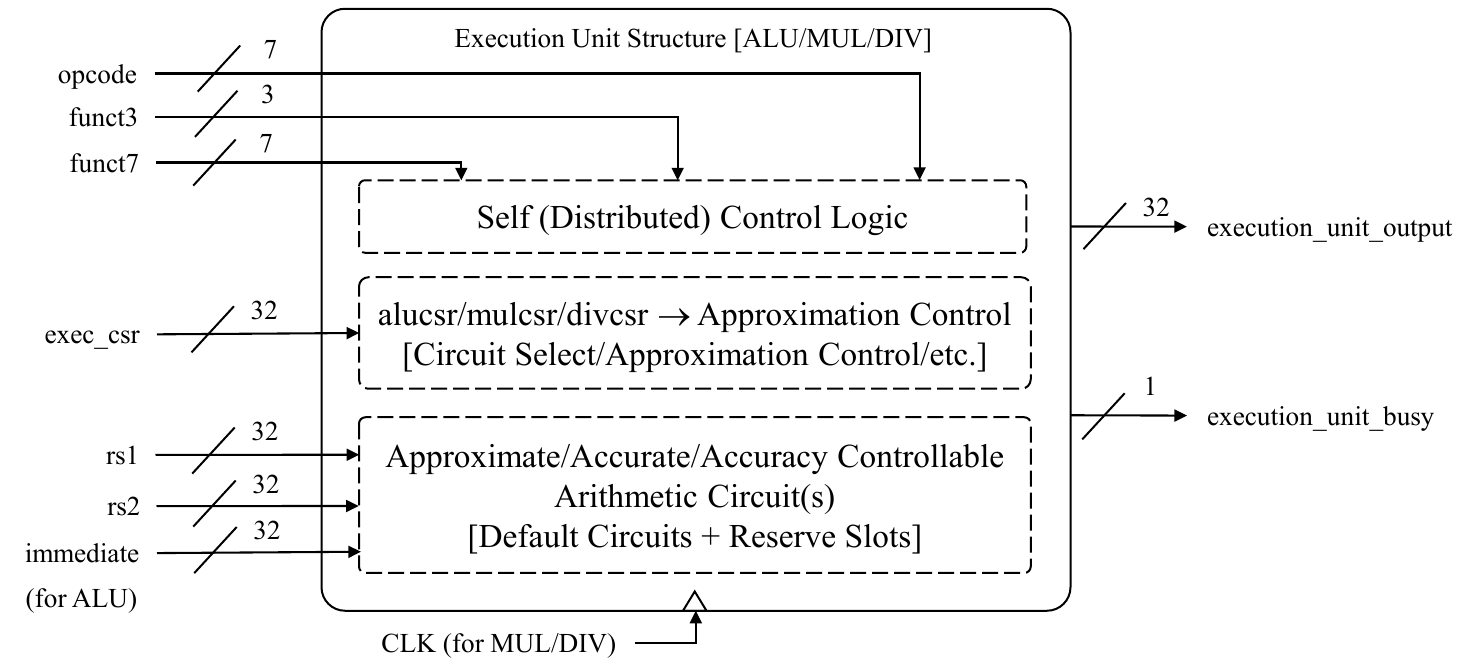}
    \caption{Proposed processor's execution units structure}
    \label{execution-units-figure}
    \end{figure}

    Each module in the execution stage consists of three primary components: the ``Self-Control Logic'', ``Approximation Control Logic'' and the ``Operational Logic''. These components work in an organized flow inside the module to ensure the correct execution of instructions within the processor. The control logic uses the decoded fields coming from decode stage to create control signals in a distributed control manner. After the generation of control signals, operations will be taken in the modules according to the instruction fields. In this architecture, designers can make changes directly in the operational logic without any need to make changes in the self-control and approximation control part, which is a beneficial feature in reconfigurable architectures. In the traditional method of CPU design, a centralized control unit is responsible for generation and propagation of control signals and addition or replacement of execution units by designers would require a modification in the control unit. The structure of execution units and their three main mentioned sections, Control Logic, Approximation Control and Operational Logic can be seen in Fig.~\ref{execution-units-figure}.

\begin{figure}[htbp]
    \centering
    \includegraphics[width=\linewidth]{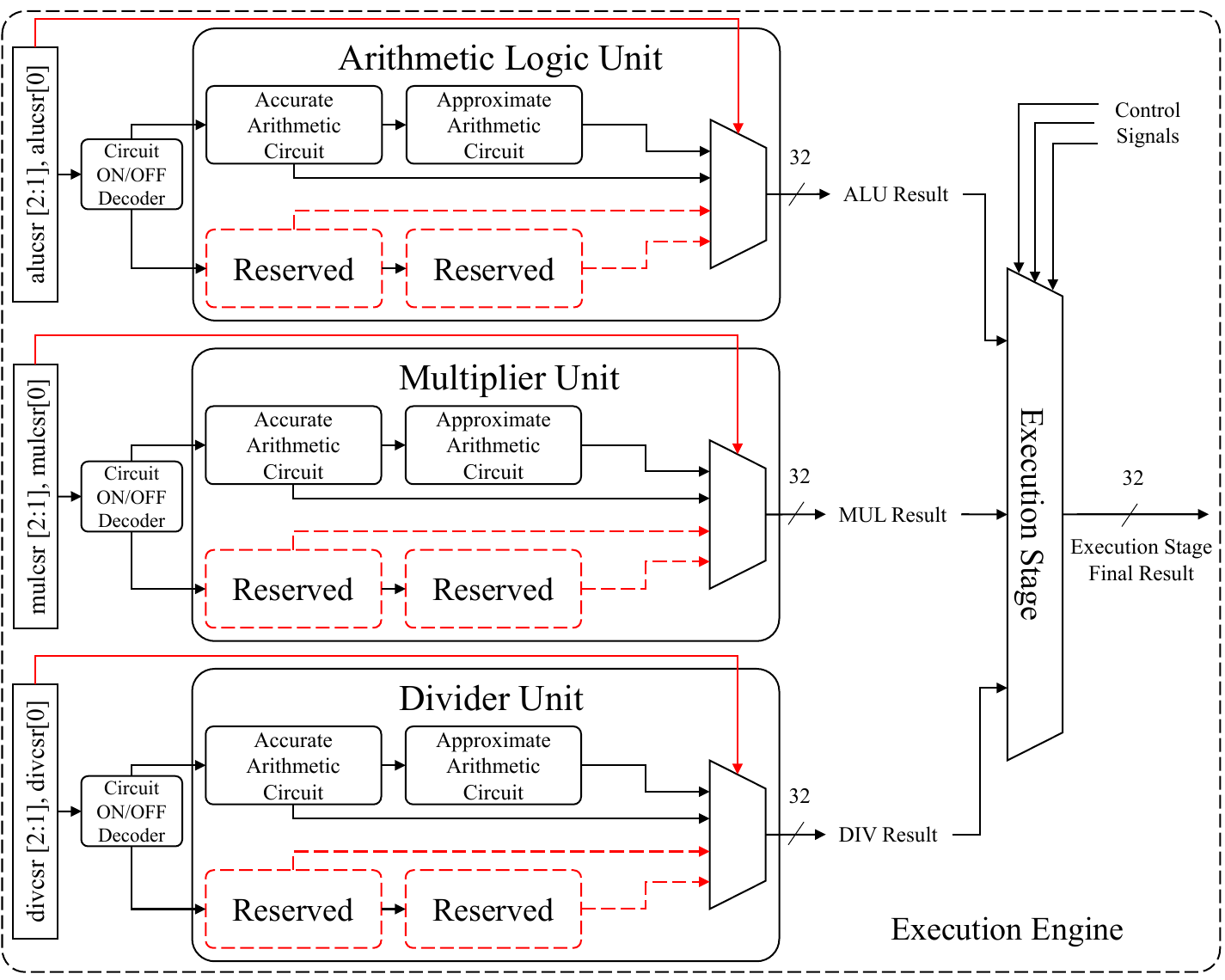}
    \caption{Circuit result selection logic in core’s execution units according to the respective CSR of each module}
    \label{circuit-selection-figure}
    \end{figure}

    Predefined approximate/accurate arithmetic circuit designs are included in the execution units, but top modules (Arithmetic Logic Unit, Multiplier Unit, Divider Unit) are designed in a way that designers can easily include and integrate their own design within the execution units as well. These execution units have reserved blocks for four modules which can be accurate or approximate, and if they are approximate, accuracy control logic is supported and handled by execution unit through the RISC-V CSRs. Our design combines the precision of accurate circuits with the efficiency and potential energy savings offered by approximate circuits. The design gains benefits of having approximate arithmetic units alongside exact computation modules capabilities in one standard core. As shown in Fig.~\ref{circuit-selection-figure}., the execution unit result will be selected by a multiplexer at the end of the execution unit hardware which is determined by the value in the relative CSR. Unused circuits are dynamically switched off by leveraging the circuit selection field in the CSRs. This helps to prevent redundant processing of input data, eliminating unnecessary dynamic and static power consumption by the unused circuits within the core which will lead to higher power-efficiency. This structure allows the user or supervisor to dynamically switch between circuits during runtime according to power consumption constraints (e.g., in battery operated embedded devices.). The key contribution of the implemented design is enabling dynamic energy control utilizing controllable error levels and approximation, where runtime energy control can be managed by user or the supervisor privilege access, in fault-tolerant applications where approximation methods are allowed in computation process.

\begin{figure}[htbp]
    \centering
    \includegraphics[width=\linewidth]{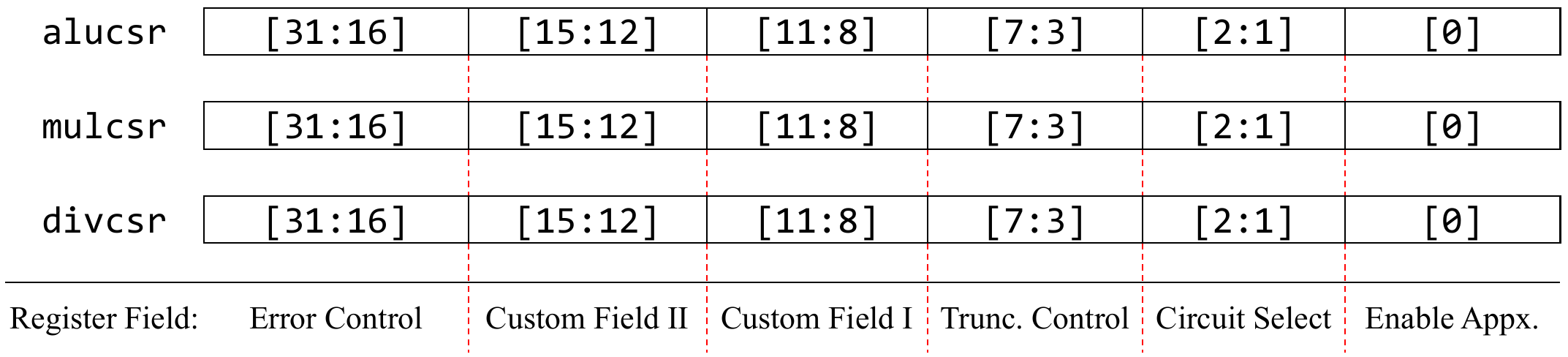}
    \caption{Proposed Platform’s Execution Engine Control Status Registers General Format}
    \label{csr-format-figure}
    \vspace{-2mm}
    \end{figure}

    The execution engine has three main modules: Arithmetic Logic Unit, Multiplier Unit and Divider Unit. These modules are implemented in a way which gives designers the ability to change or add execution circuits in the function units, without any need to change the control logic of the modules. For each execution unit, there is a dedicated special purpose register named \verb|alucsr|, \verb|mulcsr| and \verb|divcsr|. These control status registers (CSRs) are designed in a way to provide the core’s special features for approximate computing. The structure of the fields within the mentioned registers are shown in Fig.~\ref{csr-format-figure}.

    Each entry in these registers is related to a specific feature, aiming to assist users and designers in demonstrating advancements in their respective fields of study and work. The ultimate objective of these features, which contributes to the overall goal of the project, is to enhance user accessibility in programming and provide designers with greater flexibility in implementations. Definition of each field and bits in the mentioned special purpose registers are as follows:
    \begin{itemize}
    
        \item Bit 0: Enabling approximate computations.
        
        \item Bits 1 to 2: Gives the programmer the ability of selection from four arithmetic circuits in the execution unit which can be accurate or approximate as shown in Fig.~\ref{circuit-selection-figure}. One notable advantage of this feature is the design philosophy behind this project, which ensures that control logic, forwarding bases, and pipeline signals remain unchanged even in the presence of varying timing and clock cycles in each circuit.
        
        \item Bits 3 to 7: Custom field 1 for designer’s need of control and accessibility features.
        
        \item Bits 8 to 11: Custom field 2 for designer’s need of control and accessibility features.
        
        \item Bits 12 to 15: Truncation control level for special approximate arithmetic units which may work with truncation logic.

        \item Bits 16 to 31: Error control bits for accuracy controllable and error configurable approximate designs, which also include this platform’s default execution units.
    
    \end{itemize}

\begin{figure}[htbp]
    \centering
    \includegraphics[width=\linewidth]{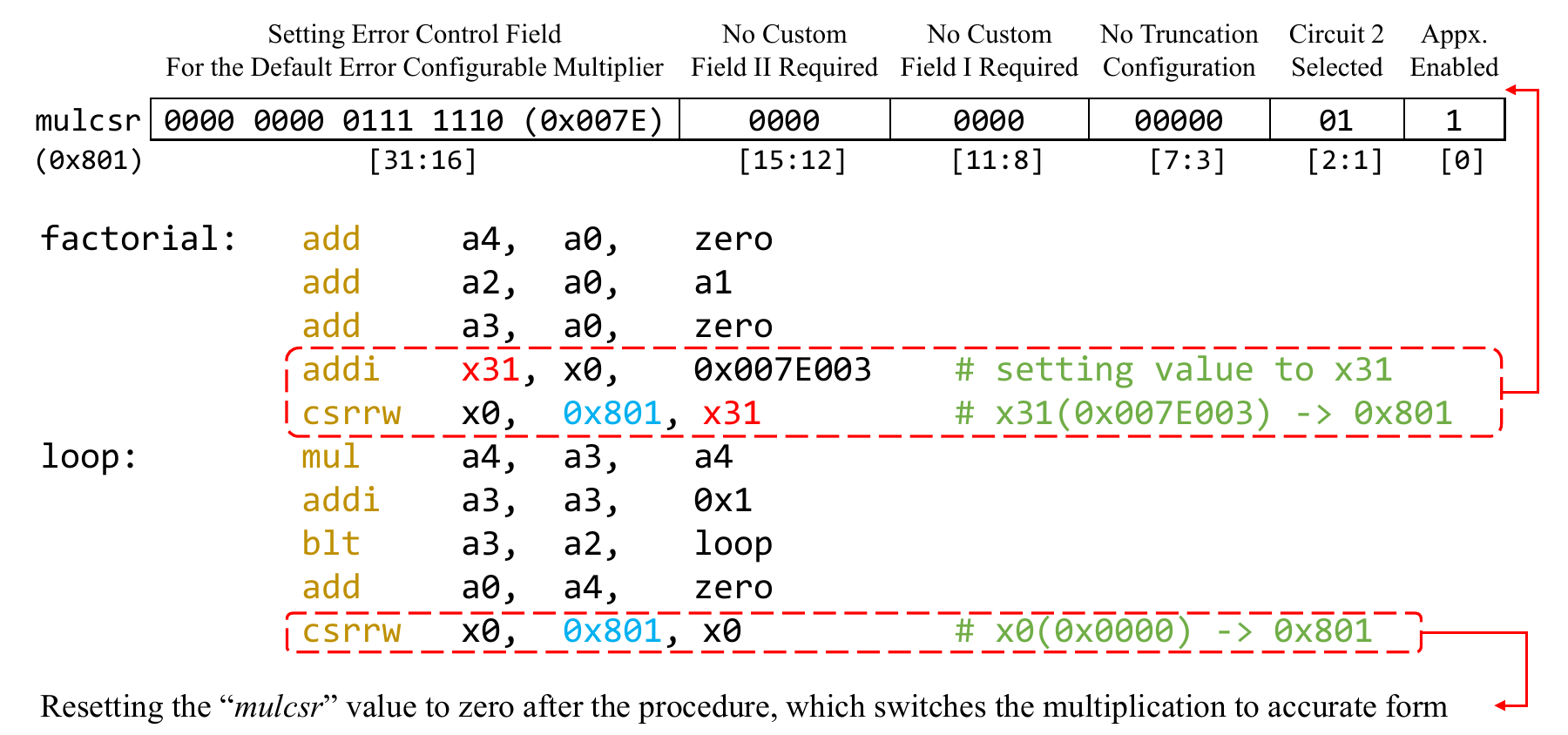}
    \caption{Sample “Factorial” RISC-V assembly code utilizing \textit{mulcsr} to select the approximate circuit and setting error level}
    \label{factorial-assembly-code}
    \end{figure}

    These three special purpose registers are mapped as 0x800 (\verb|alucsr|), 0x801(\verb|mulcsr|) and 0x802 (\verb|divcsr|) in the core’s CSR indexing range \cite{RISC-V-instruction-set-manual—volume-II}. Fig.~\ref{factorial-assembly-code} presents a sample RISC-V assembly code to show the programming convention of the platform using an approximate arithmetic circuit where the error level is configurable in the circuit. The accurate result of the procedure with input 10, is 3628800. Using the approximation factor set in the \verb|mulcsr| in Fig.\ref{factorial-assembly-code}. the processor results 3587840, which has an error of 1.128\%.
    
    One significant feature of the following format of programming convention in this platform is that unlike many other proposed designs, there is no need to add any instruction to RISC-V original compiler tool-chain \cite{gcc}. Unlike \cite{Approxrisc} and \cite{axe} approaches which was done by adding custom instructions to the original compiler, no additional modifications in software toolchain is required in the proposed method. This also prevents recompilation of existing source code for approximate execution on this platform.

\section{Execution Engine}\label{execution-engine-section}
    This processor supports RV32I(E)M instruction set of RISC-V standard extensions. Both I and E extensions are designed for integer arithmetical and logical instructions \cite{RISC-V-ISA-Extensions-A-Survey} which are configurable in this core, as designer may set the input parameter of the module in the RTL source.

    \subsection{Adder/Subtractor}
    For the main adder/subtractor circuit (demo circuit) of the core, we have implemented a fast and low power carry select adder, with reduced critical path. The proposed 32-bit adder circuit, is an error-configurable design with 8-bit error control signals (64 configurations) which will suite this module with various applications. The CSA architecture proposed in this platform is constructed by 4-bit ripple carry adders which are error controllable, and the error configuration feature in the adder is concluded from the inner ripple carry adders. This adder can also perform accurate addition when the respected field of \verb|alucsr| is equal to 0x0F, or the approximate enabling bit turned to zero. The accurate/approximate computation feature in the default adder aids the design with saving power and area, leaving the core with no need of additional arithmetic circuits supporting accurate and approximate addition exclusively. Proposed adder has improved area by 11.3\%, power by 21.1\% (regarding average power consumption in all 64 configurations which is 144.92µW) and critical path by 34.1\% in comparison with an accurate CSA. Detailed hardware efficiency metrics for the adder is reported in Table \ref{adder-hardware}, where ACC stands for conventional accurate carry select adder, and APX stands out for the proposed accurate/approximate adder. The AVG term indicates the average of all possible configurations which in this case, are 64 error levels of the proposed adder.
    
    \begin{table}[htbp]
    \caption{Hardware Efficiency Comparison of Proposed Adder and Conventional Carry Select Adder in 45nm CMOS}
    \begin{center}
    \noindent\begin{minipage}{\linewidth}
    \centering
    \begin{tblr}{
      width = \linewidth,
      colspec = {Q[85]Q[85]Q[105]Q[225]Q[85]Q[85]Q[85]Q[85]},
      cells = {c},
      cell{1}{1} = {c=8}{0.9\linewidth},
      cell{2}{1} = {c=2}{0.150\linewidth},
      cell{2}{3} = {c=4}{0.350\linewidth},
      cell{2}{7} = {c=2}{0.150\linewidth},
      hlines,
      vlines,
    }
    32-bit Carry Select Adder / 8-bit Error Control &  &  &  &  &  &  & \\
    Area (µm²) &  & Power (µW) &  &  &  & Delay (ns) & \\
    ACC & APX & ACC & APX (AVG) & Min & Max & ACC & APX\\
    337.3 & 299.2 & 183.62 & 144.92 & 136.1 & 152.0 & 2.02 & 1.33
    \end{tblr}
    \end{minipage}
    \label{adder-hardware}
    \end{center}
    \end{table}  
    
    This circuit is specifically designed for addition and subtraction instructions of I and E extension of RISC-V instruction set architecture. For address generation and control flow instructions, precise addition using an accurate adder is essential as approximation is not permissible.

    \subsection{Multiplier}
    The second default (demo) circuit integrated in the core is a low-power, high-speed error configurable approximate 8-bit multiplier which features reduced area, low energy and power consumption, and optimized critical path delay. The proposed multiplier in an 8-bit configuration showcases an improvement of 54.4\% in area occupation, 66\% in power consumption (considering the average power consumption in all dynamic configurations which is 75.49µW) and 51.5\% in critical path delay, in comparison with a conventional accurate Wallace multiplier in which details are shown in Table \ref{multiplier-hardware}. The lowest power consumption in the dynamic configuration can be achieved by the highest error level (error field of CSR equal to 0x00) with 67.43µW and the highest power is related to the highest quality in terms of error metrics (error field of CSR equal to 0x7E) with 81.05µW.
    
    \begin{table}[htbp]
    \caption{Hardware Efficiency Comparison of 8-bit Proposed Multiplier and Conventional Wallace Tree Multiplier in 45nm CMOS}
    \begin{center}
    \noindent\begin{minipage}{\linewidth}
    \centering
    \begin{tblr}{
      width = \linewidth,
      colspec = {Q[85]Q[85]Q[105]Q[225]Q[85]Q[85]Q[85]Q[85]},
      cells = {c},
      cell{1}{1} = {c=8}{0.9\linewidth},
      cell{2}{1} = {c=2}{0.150\linewidth},
      cell{2}{3} = {c=4}{0.350\linewidth},
      cell{2}{7} = {c=2}{0.150\linewidth},
      hlines,
      vlines,
    }
    8-bit Error Controllable Approximate Multiplier &  &  &  &  &  &  & \\
    Area (µm²) &  & Power (µW) &  &  &  & Delay (ns) & \\
    ACC & APX & ACC & APX (AVG) & Min & Max & ACC & APX\\
    434.7 & 198.7 & 224.31~ & 75.49 & 67.43 & 81.05 & 1.32 & 0.64
    \end{tblr}
    \end{minipage}
    \vspace{-2mm}
    \label{multiplier-hardware}
    \end{center}
    \end{table}    
    
\begin{figure}[htbp]
    \centering
    \includegraphics[width=0.8\linewidth]{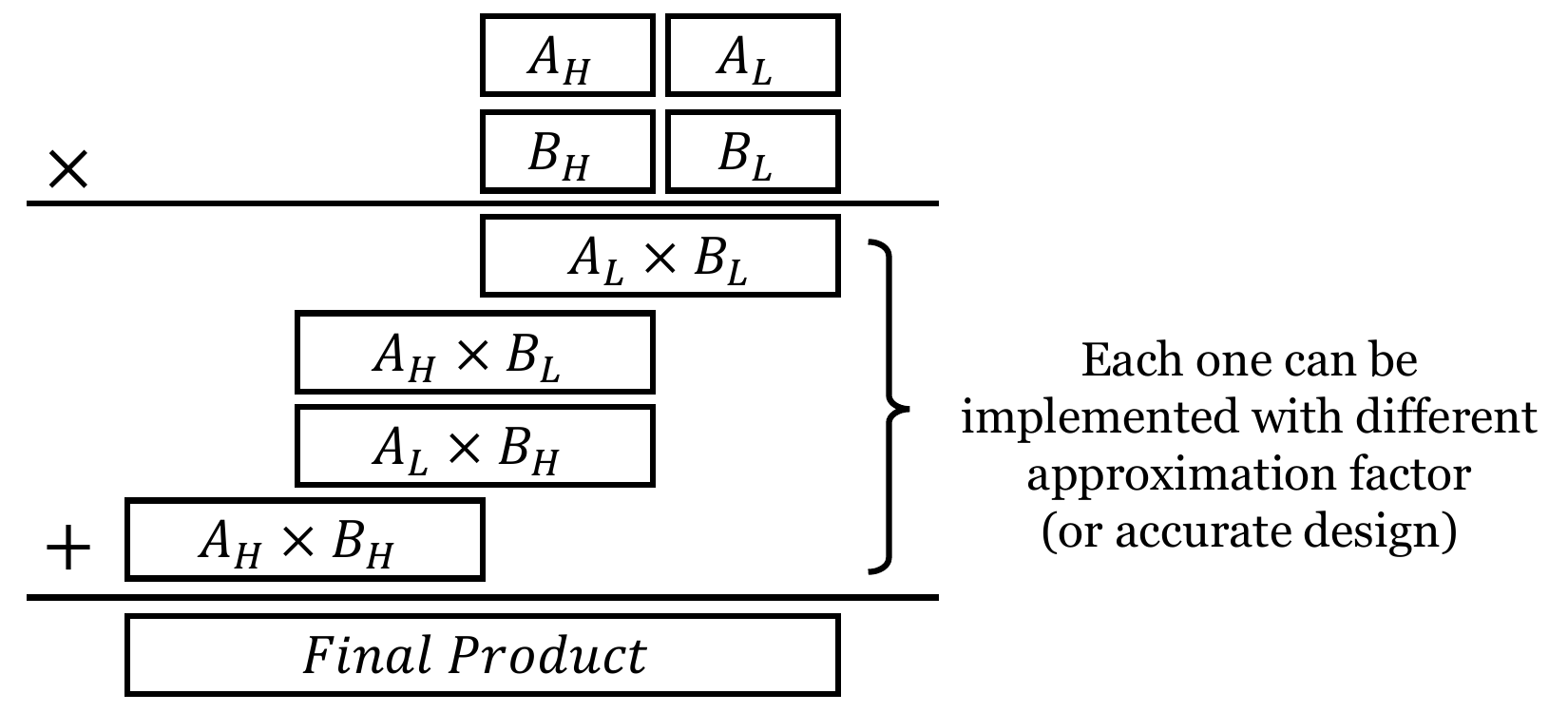}
    \caption{Hierarchical 32-bit multiplication from original 8-bit multipliers}
    \label{multiplication_32_from_8}
    \vspace{-1mm}
    \end{figure}

    For integration with the core, a 32-bit multiplier in hierarchical arrangement of 8-bit modules is implemented. The method used for creation of the 32-bit architecture uses the original 8-bit multiplier over multiple cycles to perform 16-bit multiplication. This hardware is replicated four times as illustrated in Fig. \ref{multiplication_32_from_8} to perform a 32-bit multiplication. 

    The approximate multiplier in the core has significant results on Error Rate (ER), Mean Relative Error Distance (MRED) and Normalized Mean Error Distance (NMED) \cite{error-metrics-1} \cite{error-metrics-2} which are some of the favored metrics for accuracy efficiency criteria measurement in approximate arithmetic circuits, that are described in Table \ref{error-metrics}. 

    \begin{table}[htbp]
    \caption{Description and Equations for Each Error Metric}
    \begin{center}    \noindent\begin{minipage}{0.925\linewidth}
    \centering
    \begin{tblr}{
      width = \linewidth,
      colspec = {Q[75]Q[270]Q[300]},
      cells = {m,c},
      hlines,
      vlines,
    }
    Metric & Equation & Description\\
    ER & $\frac{\#R_F}{2^{2N}}$ & Ratio of the number of false results ($\#R_F$) to number of possible outputs.\\
    MRED & $\frac{1}{2^{2N}}\sum_{i=1}^{2^{2N}}\frac{|R_i^A-R_i^X|}{R_i^A}$ & Average of $RED_i$ (Relative Error Distance) for every possible result.\\
    NMED & $\frac{\frac{1}{2^{2N}}\sum_{i=1}^{2^{2N}}|R_i^A-R_i^X|}{(2^N-1)^2}$ & Average of $ED_i$ (Error Distance) divided by most considerable exect result.
    \end{tblr}
    \end{minipage}
    \vspace{-2mm}
    \label{error-metrics}
    \end{center}
    \end{table}
    
    The results for described error analysis metrics of the proposed multiplier in an 8-bit configuration are shown in Table \ref{multiplier-table}. This table only depicts selected error levels of the multiplier, which is set by the related CSR in the processor. 

    \begin{table}[htbp]
    \caption{NMED, MRED and ER for 8-bit Version of the Proposed Approximate Dynamic Error Configurable Multiplier}
    \begin{center}
    \noindent\begin{minipage}{\linewidth}
    \centering
    \begin{tblr}{
      width = 0.8\linewidth,
      colspec = {Q[100]Q[300]Q[100]Q[100]Q[100]},
      cells = {c},
      hlines,
      vlines,
    }
    Error Level & Error Control bits (Field of CSR) & NMED (\%) & MRED (\%) & ER (\%) \\
    6 & 0x7E & 0.25 & 0.85 & 36.16\\
    5 & 0x7C & 0.26 & 0.97 & 39.73\\
    4 & 0x78 & 0.28 & 1.33 & 44.33\\
    3 & 0x70 & 0.34 & 2.11 & 50.33\\
    2 & 0x60 & 0.48 & 3.59 & 57.02\\
    1 & 0x40 & 0.76 & 5.89 & 61.8\\
    0 & 0x00 & 1.25 & 8.94 & 65.07
    \end{tblr}
    \end{minipage}
    \vspace{-4mm}
    \label{multiplier-table}
    \end{center}
    \end{table}
    
    \subsection{Divider} 
    In the last phase of core’s default execution engine development, a sample non-restoring 32-bit divider with an 8-bit error configuration range is implemented within the core, in order to finalize the functional units used in the standard RISC-V M-Extension. A similar error control mechanism for the ALU and multiplier unit is also utilized for this module, which computes accurate division at the zero error level. This control signal is delivered through the specific control register (\verb|divscr|) designated for the divider unit.
    
\section{Experimental Results in Image Processing Application}
\label{application-section}
    The assessment is based on the analysis of an application in image processing, using the proposed processor and its accurate-approximate execution engine. In this research, a widely used image sharpening algorithm, using the 5×5 kernel in \cite{image-sharp-paper} was considered. The selected case for the evaluation is a 512×512 8-bit grayscale image. The focus of approximation was exclusively on the multiplication in the convolution process, with all other operations (such as addition, subtraction, and division) being accurately executed.
    
    The evaluation of processed image quality was carried out using the peak signal-to-noise ratio (PSNR) which is defined in terms of the mean squared error (MSE). In this study, both MSE and PSNR are defined according to the following equations:
    \begin{equation}\label{mse-equation}
    MSE=\frac{1}{nm}\sum_{i=0}^{n-1}\sum_{j=0}^{m-1}[X(i,j)-Y(i,j)]^2,
    \end{equation}
    
    \begin{equation}\label{psnr-eqution}
    PSNR=10\cdot \log_{10}(\frac{MAX^2}{MSE}),
    \end{equation}

    where \(X(i,j)\) and \(Y(i, j)\) represent the expected and obtained values, \(n\) and \(m\) correspond to the image dimensions, while \(MAX\) denotes 255, as the maximum value of each pixel in the 8-bit grayscale image.
    
    The highest PSNR, which indicates highest quality, was reached through core's default multiplier at 6\textsuperscript{th} error level.
    Table \ref{image-sharpening-error-levels-table} presents the PSNR results and power consumption for different error levels which are set by \textit{Error Control Field} of \verb|mulcsr|. The default multiplier error is reduced from error zero to six as shown in Table \ref{image-sharpening-error-levels-table}. On average, there is a 12\% reduction in power consumption when compared to the precise multiplier used in the same application.

    \begin{table*}[htbp]
    \caption{PSNR and Power for Image Sharpening Algorithm Using Processor’S Default Error Configurable Multiplier}
    \begin{center}
    
    \noindent\begin{minipage}{\linewidth}
    \centering
    \begin{tblr}{
      width = \linewidth,
      colspec = {Q[100]Q[100]Q[100]Q[100]Q[100]Q[100]Q[100]Q[100]Q[100]},
      cells = {c},
      hlines,
      vlines,
    }
    Parameter & Error Level 6 & Error Level 5 & Error Level 4 & Error Level 3 & Error Level 2 & Error Level 1 & Error Level 0\\
    PSNR (dB) & 46.32 & 43.99 & 41.78 & 37.51 & 31.27 & 26.52 & 21.82 \\
    Power (mW) & 4.5544 & 4.5403 & 4.5092 & 4.4247 & 4.3412 & 4.2792 & 4.2376
    \end{tblr}
    \end{minipage}
        
    \label{image-sharpening-error-levels-table}
    \end{center}
    \vspace{-5mm}
    \end{table*}

    \begin{table*}[htbp]
    \caption{PSNR Result Comparison of Different Multipliers Using Proposed Platform}
    \begin{center}
    \definecolor{Thunder}{rgb}{0.137,0.121,0.125}
    \noindent\begin{minipage}{\linewidth}
    \centering
    \begin{tblr}{
      width = \linewidth,
      colspec = {Q[120]Q[100]Q[100]Q[90]Q[90]Q[90]Q[90]Q[90]Q[90]Q[90]},
      cells = {c},
      column{1} = {fg=Thunder},
      column{5} = {fg=Thunder},
      column{6} = {fg=Thunder},
      column{7} = {fg=Thunder},
      column{8} = {fg=Thunder},
      column{9} = {fg=Thunder},
      column{10} = {fg=Thunder},
      hlines,
      vlines,
    }
    Designs

    & Error Level 6
    & Error Level 5

    & AxRM1
    & AxRM2
    & AxRM3
    
    & Ax8-1  
    & Ax8-2 \\
    Reference
    &  Default
    &  Default
    & \cite{AxRMs-paper}
    & \cite{AxRMs-paper}
    & \cite{AxRMs-paper}
    & \cite{hybrid-partial-product-paper}
    & \cite{hybrid-partial-product-paper}\\
    PSNR (dB)
    & 46.32 
    & 43.99 
    & 48.62
    & 32.31
    & 22.51
    & 53.68
    & 23.54 \\

    \end{tblr}
    \end{minipage}
    
    \label{image-sharpening-multipliers-comparison-table}
    \end{center}
    \vspace{-5mm}
    \end{table*}
    
    Other approximate multipliers were integrated in the platform, and the same program was executed. Table \ref{image-sharpening-multipliers-comparison-table} shows a comparison between the results of this experiment. AxRMs \cite{AxRMs-paper} and Ax8 \cite{hybrid-partial-product-paper} do not benefit from dynamic reconfigurability and thus their error metrics are set at design time. And therefore, in scenarios where a trade-off between power and quality is critical, different circuits must be present in the related execution unit and the \textit{Circuit Select Field} of the \verb|csr| must be utilized. In situations where dynamic reconfigurabilty is necessary, the proposed platform's accuracy-controllable execution units provide a wide range of flexibility at execution time. 

\section{Implementation and Comparison}\label{synthesis-section}
    The proposed design has been implemented and evaluated on a Xilinx ZYNQ (XC7Z020) FPGA, and also undergone a complete ASIC synthesis and implementation flow. 
    A comparison in terms of LUT number, frequency, pipeline stages and Dhrystone \cite{dhrystone} benchmark result, between this work and similar designs, reported by \cite{fpga-paper} are shown in Table \ref{fpga-table}. On a 100MHz clock frequency, which is not the maximum frequency that phoeniX is capable of, has the highest benchmark performance, with DMIPS/MHz of 1.73.

    \begin{table}[htbp]
    \caption{LUT number, frequncy and DMIPS/MHz comparison on FPGA}
    \begin{center}

    \noindent\begin{minipage}{0.95\linewidth}
    \centering
    \begin{tblr}{
      width = \linewidth,
      colspec = {Q[160]Q[90]Q[200]Q[260]Q[202]},
      cells = {c},
      cell{1}{1} = {c=5}{0.922\linewidth},
      hlines,
      vlines,
    }
    Results
      on Xilinx 7 series FPGA &  &  &  & \\
    Design~ & LUTs & $f_{clk}$~ (MHz)~ & Pipeline
      Stages & DMIPS/MHz\\
    phoeniX & 1665 & 100 & 3 & 1.73\\
    Freedom & 2692 & 32.5 & 5 & 1.61\\
    RI5CY & 6748 & 50 & 5 & 1.1\\
    PicoRV32 & 1765 & 115 & 0 & 0.13
    \end{tblr}
    \end{minipage}
    
    \label{fpga-table}
    \end{center}
    \vspace{-3mm}
    \end{table}
    
\begin{figure}[htbp]
    \centering
    \includegraphics[width=\linewidth]{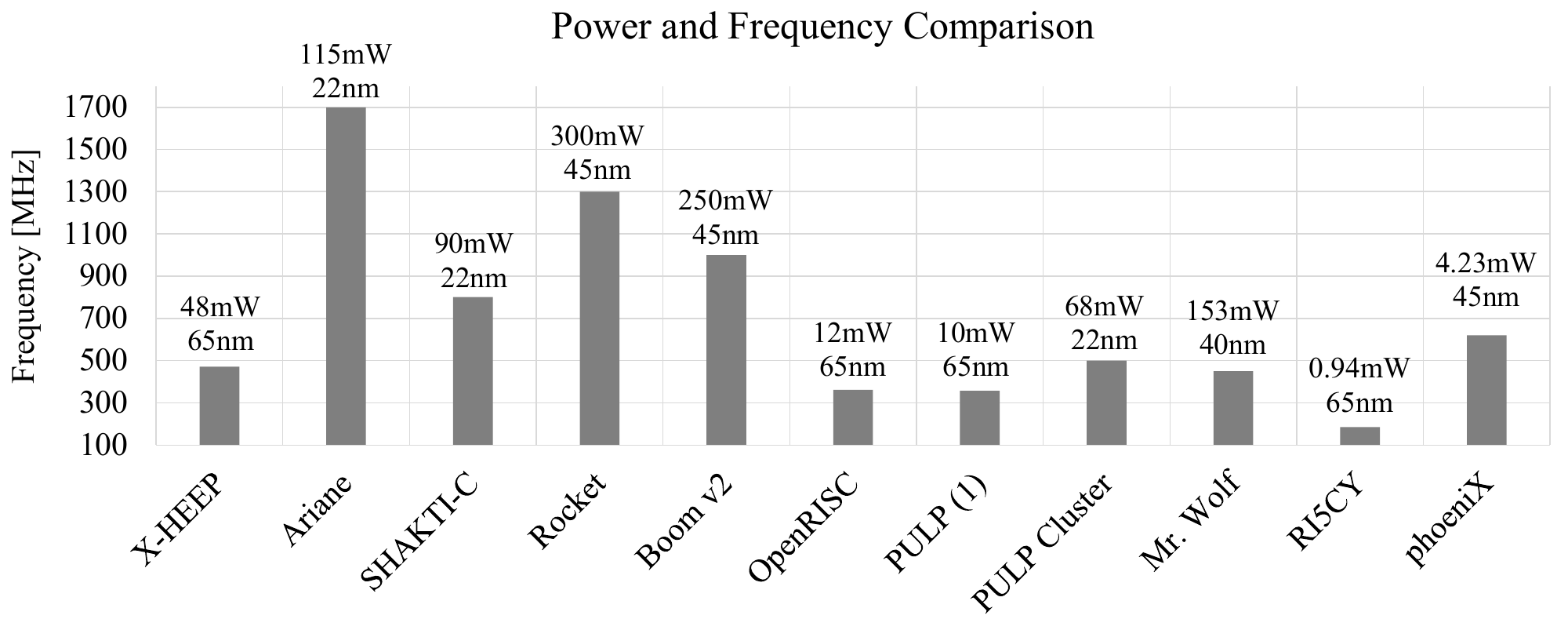}
    \vspace{-6mm}
    \caption{Power and frequency comparison between RISC-V cores}
    \label{power-comparison}
    \vspace{-1mm}
    \end{figure}

    The ASIC synthesis was done utilizing the NanGate 45nm Open Cell Library. The static timing analysis (STA) results show that the maximum delay in modules and pipeline stages is under 1.6ns. Setting the clock cycle based on the critical path provides enough margin for the maximum delay, ensuring timely data propagation through the pipeline. Adhering to this timing requirement allows the processor to operate at 620MHz, supporting efficient instruction execution and meeting operational specifications for embedded processors.
    The RISC-V Instruction Set Architecture is recognized for its high efficiency in low-power and minimal area design, attributed to its reduced instruction set and straightforward hardware implementation. 
    A notable advantage of approximate computing is energy conservation. By leveraging the strengths of both paradigms, this core has emerged as an exemplary model for energy-efficient computing while upholding an acceptable level of performance. The platform exhibits notable metrics, with an average power consumption of 4.23mW at 1.1V at 620MHz frequency, in 0.024mm² of area when using the default execution engine, showcasing its efficiency for various error-tolerant applications. 
    Fig.~\ref{power-comparison} shows a comparison in term of maximum power consumption between phoeniX and other notable RISC cores, RI5CY \cite{ri5cy-paper}, PULP (1), PULP Cluster \cite{pulp-paper-2}, X-HEEP \cite{x-heep}, Ariane \cite{ariane-paper}, SHAKTI-C \cite{shakti-2}, Rocket \cite{rocket-paper}, Boom v2 \cite{boom-paper}, OpenRISC \cite{open-risc} and Mr. Wolf \cite{mr-wolf-paper}. According to different technology nodes in these processors and by applying scaling factors, this comparison showcases remarkable power efficiency offered by the proposed core, resulted from the reconfigurable accurate/approximate execution units. Technical details about some of these cores are explained later in this section.
    
    Table \ref{pj-instructions} presents energy consumption per instruction, in sample software programs, compiled using GCC with standard optimization options for each program. Compilation options for each program executed on processor for post-synthesis power simulations are also reported in Table \ref{pj-instructions}. It is concluded from the table that the proposed platform is an energy-efficient base for embedded designs.

    \begin{table}[htbp]
    \caption{Energy Per Instruction [pJ] for Standard Sample Codes Running on the Proposed Platform}
    \begin{center}

    \noindent\begin{minipage}{0.95\linewidth}
    \centering
    \begin{tblr}{
      width = \linewidth,
      colspec = {Q[100]Q[100]Q[100]},
      row{1} = {c},
      row{2} = {c},
      row{3} = {c},
      row{4} = {c},
      row{5} = {c},
      row{6} = {c},
      row{7} = {c},
      cell{8}{1} = {c=3}{0.9\linewidth},
      cell{9}{1} = {c=3}{0.9\linewidth},
      cell{10}{1} = {c=3}{0.9\linewidth},
      cell{11}{1} = {c=3}{0.9\linewidth},
      vline{-} = {1-7}{},
      hline{1-8} = {-}{},
    }
    Application\textsuperscript{*}~ & Power
    Efficiency\textsuperscript{1} (pJ/op) & Error Control Field of  \textit{mulcsr} \\
    Dhrystone\textsuperscript{a} & 7.3423 & NULL (Accurate)\\
    BubbleSort\textsuperscript{†} & 7.8190 & 0x00\\
    Fibonacci\textsuperscript{†} & 7.8634 & 0x7E\\
    2Dconv3x3\textsuperscript{†} & 7.7609 & 0x78\\
    FindMaxArray\textsuperscript{†} & 7.9925 & 0x60\\
    FIR† & 8.1020 & 0x60\\
    \textsuperscript{1} Processor running @1.1V, 620MHz, &  & \\
    \textsuperscript{*} Compiled with riscv64-unknown-gcc 8.3.0 &  & \\
    \textsuperscript{a} Using recommended ARM compiler flags and settings in \cite{dhrystone} &  & \\
    \textsuperscript{†} GCC-8.3.0 Flags:
      -c -O2 -mabi=ilp32 -march=rv32im &  & 
    \end{tblr}
    \end{minipage}
    \vspace{-4mm}
    \label{pj-instructions}
    \end{center}
    \end{table}

    Table \ref{pj-comparison} is an energy efficiency comparison between RISC-V cores in pJ per operation metric. The proposed design with an average of 7.85 pJ/op is the most efficient design in the study with other reported values in term of energy per instruction. Mr. Wolf \cite{mr-wolf-paper} and PULP cluster \cite{pulp-paper-2} are second and third best designs, noting the fact that technology nodes differ in this comparison. Total energy consumption in each application and average result of 7.85 pJ/op was measured through post-synthesis power simulations alongside switching activity files extracted from the set of applications executed on the processor. All measurements for the proposed design were done under same circumstances (including voltage, frequency and compiler options) described in Table \ref{pj-instructions}).

    \begin{table}[htbp]
    \caption{Energy Efficiency [pJ/op] Comparison in Selected RISC-V Processors}
    \begin{center}

    \noindent\begin{minipage}{0.9\linewidth}
    \centering
    \begin{tblr}{
      width = \linewidth,
      colspec = {Q[235]Q[140]Q[320]Q[220]},
      row{odd} = {c},
      row{2} = {c},
      row{4} = {c},
      row{6} = {c},
      row{8} = {c},
      cell{10}{1} = {c=4}{0.931\linewidth},
      vline{-} = {1-9}{},
      hline{1-10} = {-}{},
    }
    RISC-V Processor & Ref. & Energy Efficiency (pJ/op) & Tech. Node (nm)\\
    Ariane & \cite{ariane-paper} & 51.8 & 22\\
    Rocket & \cite{rocket-paper}\textsuperscript{*} & 100 & 45\\
    Mr. Wolf & \cite{mr-wolf-paper}\textsuperscript{*} & 12.5 & 40\\
    Boom v2 & \cite{boom-paper}\textsuperscript{*} & 133 & 45\\
    SHAKTI & \cite{shakti-2}\textsuperscript{*} & 122 & 22\\
    PicoRV32 & \cite{pico-paper} & 92.6 & 130\\
    PULP Cluster & \cite{pulp-paper-2} & 25.2 & 65\\
    phoeniX & Proposed & 7.85 & 45\\
    \textsuperscript{*} Reported by \cite{ariane-paper} &  &  & 
    \end{tblr}
    \end{minipage}
    \vspace{-7mm}
    \label{pj-comparison}
    \end{center}
    \end{table}

    Table \ref{architecture-table} provides an architectural and technical analysis and comparison between renowned processors.
    The comparison focuses on overall performance analysis and hardware efficiency in which the proposed processor was considered with accurate execution engine in order to conduct a comparison with other processors which do not have support for approximation by default.
    In this research, processors are categorized as application class, and embedded or microcontroller class according to their architectural specialties. The proposed design is architecturally categorized as an embedded core, but as it is shown in Table \ref{architecture-table}, it has a frequency rate higher than most embedded processors, and on the edge of the transition between embedded computing and high-performance computing. The optimized path of the core which is less than 24 NAND gates, leads to a remarkable result in term of critical path and overall performance. In fact, this project represents an optimized embedded class platform with significantly reduced area and power consumption, which can perform with a frequency and performance near application class and high-performance processors.
    
    Dominating industrial cores in microcontrollers are designed by ARM, known as Cortex-M series. These processors are usually fabricated on more mature nodes such has 180nm, 90nm and 40nm technologies. ARM Cortex-M series \cite{arm-cortex-m} are architecturally close to the proposed core and other designs in the research, except for Cortex-M7 which is a superscalar 6-stage pipelined core. Cortex-A5 is an application class, single issue 8-stage pipelined in-order processor, which in matter of frequency and performance level can be compared with the presented platform.
    
    Table \ref{architecture-table} also presents a comparison within these cores (including application class and embedded class) based on result of the Dhrystone benchmark. The proposed processor has a rate of 1.89 DMIPS/MHz, Dhrystones per Seconds per MHz of 3324 and a CPI of 1.13 at the operating frequency. It is important to note that some of this cores such as Ariane, Boom v2 and Rocket are application class and have a frequency rate of +1GHz, through a tape out process with newer technology nodes. SHAKTI \cite{shakti-2} was taped-out in 22nm technology with a maximum frequency level of 800MHz and is an application class core. Ariane, designed by ETH Zurich was crafted in 22nm FDSOI technology \cite{ariane-paper} as an in-order (out-of-order execute, in-order commit) 6-stage pipelined processor. Rocket \cite{rocket-paper} by UC Berkeley, is a 6-stage pipelined design in 45nm technology. 
   
    \begin{table*}[htbp]
    \caption{Architectural and Performance Comparison of Processors With Different Technology Nodes}
    \begin{center}

    \noindent\begin{minipage}{\linewidth}
    \centering
    \begin{tblr}{
      width = \linewidth,
      colspec = {Q[85]Q[63]Q[95]Q[83]Q[95]Q[88]Q[92]Q[90]Q[215]},
      row{1} = {c},
      row{2} = {c},
      row{3} = {c},
      row{4} = {c},
      row{5} = {c},
      row{6} = {c},
      row{7} = {c},
      row{8} = {c},
      row{9} = {c},
      row{10} = {c},
      row{11} = {c},
      row{12} = {c},
      row{13} = {c},
      row{14} = {c},
      row{15} = {c},
      cell{2}{1} = {c=9}{0.93\linewidth},
      cell{8}{1} = {c=9}{0.93\linewidth},
      cell{16}{1} = {c=9}{0.93\linewidth},
      vline{-} = {1,3-7,9-15}{},
      hline{1-16} = {-}{},
    }
    Processor & Ref. & Power (mW) & Node (nm) & Freq. (MHz) & Area (mm²) & DMIPS/MHz & ISA & Architecture\\
    Application Class and High Performance &  &  &  &  &  &  &  & \\
    Ariane & \cite{ariane-paper} & \textbf{52} & 22 & \textbf{1700} & 0.30 & 1.65 & RV64IMC & 6-stage, out-of-order
      pipeline\\
    SHAKTI-C & \cite{shakti-2} & 90 & 22 & 800 & 0.29 & 1.68 & RV64G & 5-stage, in-order
      pipeline\\
    Rocket & \cite{rocket-paper} & 125 & 45 & 1300 & \textbf{0.14} & 1.72 & RV32/64GC & 6-stage, in-order
      pipeline\\
    Boom v2 & \cite{boom-paper} & 250 & 45 & 1000 & 1.00 & \textbf{2.13} & RV64G & 8-stage, out-of-order
      pipeline\\
    Cortex-A5 & \cite{arm-cortex-a} & $\geq{500}$ & 40 & 1000 & 0.53 & 1.57 & ARMv7-A & 8-stage, in-order
      pipeline\\
    Embedded and Microcontroller Class &  &  &  &  &  &  &  & \\
    OpenRISC & \cite{open-risc} & 12 & 65 & 362 & 0.064 & NR* & RISC (Org.) & 4-stage, in-order
      pipeline\\
    RI5CY & \cite{ri5cy-paper} & 3.77 & 65 & 560 & 0.059 & 1.10 & RV32IM & 4-stage, in-order pipeline\\
    CV32E40P & \cite{ibex-paper} & \textbf{1.57} & 65 & 453 & 0.104 & NR* & RV32IMC & 4-stage, in-order
      pipeline\\
    Ibex & \cite{ibex-paper} & 1.78 & 65 & 394 & 0.053 & 0.90 & RV32IMC & 2-stage, in-order
      pipeline\\
    Cortex-M4 & \cite{arm-cortex-m} & 7.87 & 40 & 240 & 0.028 & 1.67 & Armv7E-M & 3-stage, in-order
      pipeline\\
    PicoRV32 & \cite{pico-paper} & 5.14 & 130 & 250 & 0.23 & 0.50 & RV32IMC & Multi-cycle RISC\\
    phoeniX & Proposed & 4.23 & 45 & \textbf{620} & \textbf{0.024} & \textbf{1.89} & RV32IEM & 3-stage, in-order
      pipeline\\
    NR*: Not Reported &  &  &  &  &  &  &  & 
    \end{tblr}
    \end{minipage}

    \label{architecture-table}
    \end{center}
    \vspace{-7mm}
    \end{table*}
    
    In application class processors Ariane stands out as a low power design, powered by the 22nm node, with highest frequency rate of 1.7GHz. Rocket is the smallest core in this class with an area of 0.14mm² in 45nm technology. Boom has a benchmark result of 2.13 DMIPS/MHz due to the out of order execution design, showcasing best application performance in the benchmarking in its class.
    RI5CY \cite{ri5cy-paper} by PULP Platform and CV32E40P \cite{ibex-paper} by OpenHW Group, are embedded cores with similar microarchitecture to phoeniX, both implemented in 65nm CMOS technology with a 4-stage in-order pipeline design. Ibex\cite{ibex-paper} by lowRISC, inspired from Zero-riscy \cite{ri5cy-paper} is an optimized 2-stage in-order pipeline core, implemented in 65nm CMOS, with a configurable RV32IMC architecture, resulting a DMIPS/MHz rate of 0.9. 

    In conclusion, phoeniX stands out in benchmark result, with 1.89 DMIPS/MHz, and frequency rate of 620MHz, as the best in each category in embedded cores. The proposed processor also has the smallest area in comparison with other designs in this study, occupying only 0.024mm² in 45nm technology. CV32E40P has the lowest power consumption in 453MHz frequency. Ibex as a small 2-stage pipelined core is the second best in term of power consumption with 1.78mW. The pheoniX core is the third in this factor, with average power consumption of 4.23mW in 620MHz frequency, in 45nm CMOS technology, regarded as low-power and energy-efficient platform. 
    
\section{Conclusion}\label{conclusion-section}

    In conclusion, approximate computing holds significance as it enables quicker and more energy-efficient calculations by trading off a small amount of accuracy for performance gains. This strategy is especially beneficial in scenarios where precise outcomes are not always crucial, like in image processing or machine learning tasks. On the other hand, presence of low power processors in embedded systems is essential to extend and support functionality in environments with limited energy resources.
    The importance and necessity of a reconfigurable platform for implementing approximate circuits are concluded in these statements, aiming to achieve energy efficiency and performance enhancements while preserving application-level accuracies in fault-tolerant applications.

    This paper introduces a reconfigurable embedded platform, proposing a novel way of utilizing approximate computation in a RISC-V core, which is the key goal of the project. The project’s focus on modularity and the removal of traditional control unit, sets it apart from many other designs, providing an ideal base for computer architecture, digital design and fault-tolerant application researches.

    The platform meets the need for energy-efficient and high-performance embedded systems, through employing energy saving by approximation, while maintaining acceptable application-level accuracy. This platform enables integration of approximate arithmetic units with diverse structures and varying levels of accuracy within the core, by complying with signaling conventions of the platform. This allows for customizable trade-offs between speed, accuracy, and power consumption, based on the specific needs of the application. Additionally, its architecture supports the easy integration of additional units such as hardware accelerators, enhancing task-specific performance.
    
    In further and future works and updates, phoeniX will be integrated into an NoC (Network on Chip) in order to be a part of an MPSoC. 
    The proposed design gives the flexibility of dynamic accuracy and error level control in one core, which means the integration of the processor in an MPSoC platform can give us the benefits of both homogeneous and heterogeneous systems in various scenarios and applications.

\section*{Acknowledgment}

The authors are grateful to A. M. H. Monazzah for their valuable insights, which greatly improved this paper.

\end{document}